\documentclass{article}
\usepackage{epsfig}
\newcommand{\cosec}{{\rm cosec}}
\newcommand{\sech}{{\rm sech}}
\newcommand{\cosech}{{\rm cosech}}

\def\eq{\!\!\!\! &=& \!\!\!\! }
\def\Ad{A^{\dagger}}
\def\nn{\nonumber}
\def\nonu{\nonumber}
\def\br{\begin{eqnarray}}
\def\er{\end{eqnarray}}
\def\be{\begin{equation}}
\def\ee{\end{equation}}

\begin{document}
~\hfill {\footnotesize UICHEP-TH/98-11}\\

\begin{center}
\Large{\bf Exactly solvable models of supersymmetric quantum
mechanics and connection to spectrum generating algebra}
\end{center}

\vspace*{.25in}

\begin{center}\large{
  Asim Gangopadhyaya$^{a,}$\footnote{e-mail: agangop@luc.edu, asim@uic.edu}~,
  Jeffry V. Mallow $^{a,}$\footnote{e-mail: jmallow@luc.edu}~,
  Constantin Rasinariu $^{b,}$\footnote{e-mail: costel@uic.edu}~ \\
  and  Uday P. Sukhatme$^{b,}$\footnote{e-mail: sukhatme@uic.edu}
}
\end{center}
\date{}

\begin{center}
    \begin{tabular}{l l}
$a$) &Department of Physics, Loyola University Chicago, Chicago, USA \\
$b$) &Department of Physics, University of Illinois at Chicago,
     Chicago, USA
    \end{tabular}
\end{center}

\vspace{0.15in}

\begin{abstract}

For nonrelativistic Hamiltonians which are shape invariant, analytic expressions
 for the eigenvalues and eigenvectors can be derived using the well known method
 of supersymmetric quantum mechanics. Most of these Hamiltonians also possess
 spectrum generating algebras and are hence solvable by an independent group
 theoretic method. In this paper, we demonstrate the equivalence of the two
 methods of solution by developing an algebraic framework for shape invariant
 Hamiltonians with a general change of parameters, which involves nonlinear
 extensions of Lie algebras.

\end{abstract}

\section{Introduction}

Supersymmetric quantum mechanics (SUSYQM) \cite{Cooper} provides an elegant and
 useful prescription for obtaining closed analytic expressions both for the
 energy eigenvalues and eigenfunctions of a large class of one dimensional
 problems. The main ingredients in SUSYQM are the supersymmetric partner
 Hamiltonians $H_{-}\equiv \Ad A$ and $H_{+} \equiv A \Ad$. The $A$ and $\Ad$
 operators used in this factorization are expressed in terms of the
 superpotential $W$ as follows:
\begin{equation}
\label{A}
   A(x,a) = \frac{d}{dx} + W(x,a) \quad;\quad \Ad (x,a) =
-\frac{d}{dx} + W(x,a)~.
\end{equation}
Here, $W$ is a real function of $x$ and $a$ is a parameter (or a set of
 parameters), which plays an important role in the approach of this paper. An
 interesting feature of
SUSYQM is that for a shape invariant system \cite{Infeld}, i.e. a system
 satisfying an integrability condition
\begin{equation}
\label{si}
   W^2(x,a_0) + \frac{dW(x,a_0)}{dx} = W^2(x,a_1) - \frac{dW(x,a_1)}{dx} +
 R(a_0)~;
   \quad a_1 = f(a_0)~~,
\end{equation}
the entire spectrum can be determined algebraically without ever referring to
 underlying differential equations \cite{Cooper}.

Several of these exactly solvable systems are also known to possess what is
 generally referred to as a spectrum generating algebra (SGA)
 \cite{Alhassid,SGA}. The Hamiltonian of these systems can be written as a
 linear or quadratic function of an underlying algebra, and all the quantum
 states of these systems can be determined by group theoretic methods.

One may naturally ask the question whether there
is any connection between a general shape invariance condition and a spectrum
generating algebra. In this paper we address the equivalence between the two
approaches, considering a large class of change of parameters, including
translations, scalings, projective transformations, as well as more complicated
functions $f(a_0)$.

In sec. 2,  we start with a general shape invariant model. We
make use of the operators $A$ and $\Ad$ to construct a three generator algebra.
 In particular, the shape
invariance condition plays a crucial role in closing the algebra, which turns
 out
to be either $so(2,1)$ or a deformation of it.  In sec. 3, several examples are
presented. In particular, we discuss shape invariant potentials generated by a
 change of parameters corresponding to translation $a_1=a_0+k$ and pure scaling
 $a_1=q a_0, q= {\rm constant}~(0<q<1)$. For the case of scaling, we find that
 the associated potential algebra is a nonlinear deformation of $su(2)$. We
also describe potential algebraic structure of cyclic potentials
\cite{Sukhatme} described as a series of shape invariant potentials which
 repeats after a cycle of $k$ iterations.  And finally, we discuss the potential
 algebra of Natanzon potentials \cite{Natanzon} and show that all translational
 shape invariant potentials can be generated from them.

\section{The Algebraic Shape Invariant Model}
To begin the construction of the operator algebra, let us express
the shape invariance condition
eq. (\ref{si}) in terms of $A$ and $\Ad$ :
\begin{equation}
\label{asi}
   A(x,a_0) \Ad (x,a_0) -  \Ad(x,a_1) A (x,a_1) =  R(a_0)~.
\end{equation}
This relation resembles a commutator structure. To obtain a closed
$su(2)$-like algebra, we introduce an auxiliary variable $\phi$ and
define the following operators
\begin{equation}
\label{asim}
    J_{+} = e^{ip\phi} \Ad(x, \chi(i\partial_\phi)) ~~,
    ~~J_{-} =  A(x, \chi(i\partial_\phi)) \: e^{-ip\phi}~,
\end{equation}
where $p$ is an arbitrary real constant and $\chi$ is an arbitrary,
real function. The operators
$A(x, \chi(i\partial_\phi))$ and $\Ad(x, \chi(i\partial_\phi))$
are obtained from eq. ({\ref{A}) with the substitution
$a_0 \rightarrow \chi(i\partial_\phi)$. This generalization is analogous to the
 familiar spherical coordinate separation of variables scheme, in which
\mbox{$\partial_\phi \Phi(\phi)\rightarrow {\rm constant}\times \Phi(\phi)$}; in this
 case the constant eigenvalue is $a_0$. From eq. (\ref{asim}), one obtains
\begin{equation}
\label{q}
[J_{+},J_{-}] =  e^{ip\phi}
    \Ad(x, \chi(i\partial_\phi))A(x, \chi(i\partial_\phi)) e^{-ip\phi}
   - A(x, \chi(i\partial_\phi))
   \Ad(x, \chi(i\partial_\phi))~.
\end{equation}
Eq. (\ref{q}) can be easily cast into the following form
\begin{equation}
\label{j+/-}
  [J_{+},J_{-}] =
  -\left\{ A(x, \chi(i\partial_\phi))\Ad(x, \chi(i\partial_\phi))
  - \Ad(x, \chi(i\partial_\phi + p))A(x, \chi(i\partial_\phi + p)) \right\}~.
\end{equation}
At this point if we judiciously choose a function  $\chi(i\partial_\phi)$ such
 that
$\chi(i\partial_\phi+p) = f[\chi(i\partial_\phi)]$, the r.h.s.
of eq.(\ref{j+/-}) can be simplified using shape invariance condition
\begin{equation}
\label{A-Ad}
  A(x, \chi(i\partial_\phi))\Ad(x, \chi(i\partial_\phi))
  - \Ad(x, \chi(i\partial_\phi + p))A(x, \chi(i\partial_\phi + p))
  = R(\chi(i\partial_\phi))~,
\end{equation}
where  we have identified
\begin{equation}
\label{a01}
  a_0 \rightarrow \chi(i\partial_\phi) \quad ; \quad
  a_1 = f(a_0) \rightarrow f[\chi(i\partial_\phi)]=\chi(i\partial_\phi+p)~.
\end{equation}
The last step in our construction is to define the operator $J_3$ as
$J_3 = -\frac{i}{p}\partial_\phi$~.
As a consequence, we obtain a deformed Lie algebra whose generators
$J_{+}, J_{-}$ and $J_3$
satisfy the commutation relations
\begin{equation}
\label{lie}
   \lbrack J_3, J_{\pm} \rbrack = \pm \;J_{\pm}~~ ;~~
   \lbrack J_{+},J_{-} \rbrack = \xi(J_3)~,
\end{equation}
where $\xi(J_3) \equiv -R(\chi(i\partial_\phi))~$ defines the deformation. Thus
 we see that shape invariance condition plays an indispensible role in the
 closing of this algebra.

Depending on the choice of the $\chi$ function in eq. (\ref{a01}), we have
different reparametrizations corresponding to several models. For example
we have
\begin{enumerate}
    \item translational models:
      $a_1=a_0+p$ for $\chi(z)=z$~(in these models if $R$ is a linear function
 of $J_3$ the algebra turns out to be $so(2,1)$ or $so(3)$
 \cite{Gangopadhyaya_proc} ; a similar conclusion was reached by
 Balantekin\cite{Balantekin} by using a somewhat different method;)
    \item scaling models:
      $a_1 = e^p a_0 \equiv q a_0$ for $\chi(z)=e^z$~,
    \item cyclic models:
      $a_1 = \frac{\alpha a_0 + \beta}{\gamma a_0 +\delta}$~, for
      $\chi(z) =
 \frac{(\lambda_1-\delta)\lambda_1^{z/p}+(\lambda_2-\delta)\lambda_2^{z/p}B(z)}
      { \gamma\left[\lambda_1^{z/p}+\lambda_2^{z/p}B(z)\right]}~$~,
\end{enumerate}
where  $\lambda_{1,2}$ are solutions of the equation
$(x-\alpha)(x-\delta)-\beta\gamma = 0$ and $B(z)$  is an arbitrary periodic
 function of $z$ with  period $p$.

Other changes of parameters
follow from more complicated choices for $\chi(z)$. For example, if one
takes $\chi(z)=e^{e^z}$, one gets the change of parameters $a_1=a_0^2$.

Note that the quantity $J_+J_-$ corresponds to the Hamiltonian
\begin{equation}
\label{H_-=AA^d}
     H_-(x,i\partial_\phi+p)= \Ad(x, \chi(i\partial_\phi + p))
     A(x, \chi(i\partial_\phi + p))~.
\end{equation}
To find the energy spectrum of the Hamiltonian $H_{-}$ of eq. (\ref{H_-=AA^d}),
 we
first construct the unitary representations of the deformed Lie algebra
defined by eqs. (\ref{lie}). The technique proceeds as follows \cite{Rocek}.
Define, up to an additive constant, a function
$g(J_3)$ such that
\begin{equation}
\label{g-def}
   \xi(J_3) = g(J_3)-g(J_3 - 1) ~.
\end{equation}
The Casimir of this algebra is then given by $C_2=J_{-} J_{+} + g(J_3)$.
It is known that in a basis in which $J_3$ and $C_2$ are diagonal, $J_{+}$
and $J_{-}$ play the role of raising and lowering operators, respectively.
Operating on an arbitrary state $\vert h \rangle $ we have
\begin{eqnarray}
\label{state}
  J_3 \vert h\rangle \eq h \vert h\rangle~, \nonumber \\
  J_{-} \vert h\rangle \eq a(h) \; \vert h-1\rangle~, \nonumber \\
  J_{+} \vert h\rangle \eq a^{\star}(h+1)\; \vert h+1\rangle ~.
\end{eqnarray}
Using eqs. (\ref{lie}) and (\ref{state}) we obtain
\begin{equation}
\label{a}
\vert a(h) \vert^2 - \vert a(h+1) \vert^2 = g(h) - g(h-1) ~.
\end{equation}
The profile of $g(h)$ determines the dimension of the unitary representation.
For example, let us consider the two cases
presented in fig. \ref{g-plot}.
\begin{figure}[ht]
\centering
   \epsfig{file=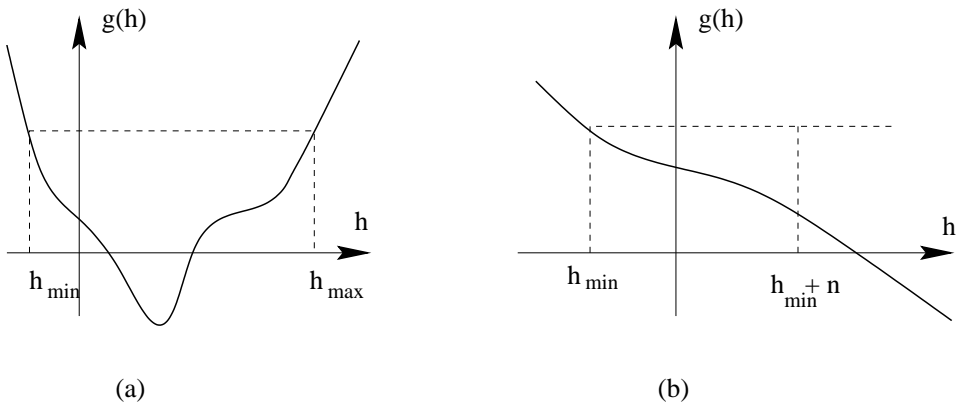, width=5in}
   \caption{Generic behaviors of $g(h)$.}
   \label{g-plot}
\end{figure}
One obtains finite dimensional representations \mbox{fig. 1a}, by starting
from a point on the \mbox{$g(h)$ vs. $h$} graph corresponding
to $h=h_{min}$, and moving in integer steps parallel to the $h$-axis till the
point corresponding to $h=h_{max}$. At the end points,
$a(h_{min})=a(h_{max}+1)=0$, and we get a finite representation.
(This is the case of $su(2)$ for example, where $g(h)$ is
given by the parabola $h(h+1)$.) If $g(h)$ is decreasing monotonically,
fig. 1b, there exists only one end point at $h=h_{min}$. Starting from $h_{min}$
 the
value of $h$ can be increased in integer steps till infinity.
In this case we have an infinite dimensional representation.
As in the finite case, $h_{min}$ labels the representation.
The difference is that here $h_{min}$ takes continuous values.
Similar arguments apply for a monotonically increasing function $g(h)$.

Having the representation of the algebra associated with a characteristic
model, we obtain (using eq. (\ref{H_-=AA^d},\ref{a}) ) the complete spectrum of
the system.
To illustrate how this mechanism works, we investigate few examples
in the next section.

\section{Examples}

\subsection{Self-Similar Potentials}
The first example is for a scaling change of parameters $a_1=qa_0$. Consider the
 simple
choice $R(a_0)=r_1a_0$, where $r_1$ is a constant. This choice
generates self-similar potentials studied in refs. \cite{Shabat1,Barclay}.
In this case, eqs. (\ref{lie}) become:
\begin{equation}
\label{scaling}
   \lbrack J_3, J_{\pm} \rbrack = \pm \;J_{\pm} \quad ; \quad
   \lbrack J_{+},J_{-} \rbrack = \xi(J_3) \equiv  -r_1~\exp(-pJ_3) ~,
\end{equation}
which is a deformation of the standard $so(2,1)$ Lie algebra.

For this case, from eqs. (\ref{scaling}) and (\ref{g-def}) one gets
\begin{equation}
\label{g}
    g(h) = \frac{r_1}{e^p - 1} e^{-p h} = -\frac{r_1}{1-q} q^{- h}
    \quad;\quad q=e^p~.
\end{equation}
Note that for scaling
problems \cite{Barclay}, one requires $0<q<1$, which leads to $p<0$.
From the monotonically decreasing profile of the function $g(h)$, it follows
that the unitary representations of this algebra are infinite dimensional.
If we label the lowest weight state of the operator $J_3$ by $h_{min}$,
then $a(h_{min})=0$. Without loss of generality we can choose the
coefficients $a(h)$ to be real. Then one obtains from (\ref{a}) for an
arbitrary \mbox{$h = h_{min} + n,~ n=0,1,2,{\ldots} $}
\begin{equation}
\label{aj}
  a^2(h) = g(h - n - 1) - g(h - 1) =
r_1~ \frac{q^n-1}{q-1} \: q^{1-h}~.
\end{equation}
The spectrum of the Hamiltonian $H_-(x,a_1)$ is given by
\begin{equation}
\label{spectrum}
H_-\vert h \rangle = a^2(h) \vert h \rangle = r_1~ \frac{q^n-1}{q-1} \: q^{1 -
 h}\vert h \rangle~.
\end{equation}
Therefore, the eigenenergies are
\begin{equation}
\label{energy}
E_n(h) = ~r_1 \alpha(h) \frac{q^n-1}{q-1} \quad ;\quad \alpha(h) \equiv q^{1-h}
 ~.
\end{equation}
To compare the above spectrum obtained using a group theoretic method with the
 results obtained from SUSYQM \cite{Barclay}, we go to
the $x$-representation. Here
$\vert h \rangle \propto e^{ip h\phi}\psi_{h_{min},n}(x)$
and hence, the Schr\"odinger equation for the Hamiltonian $H_-$ reads
$$
\left\{ -\frac{d^2}{dx^2}
      + W^2(x,e^{i\partial_\phi +p})
      - W'(x,e^{i\partial_\phi +p}) -E \right\}
       e^{i p \phi h}\psi_{h_{min},n}(x) = 0 ~,
$$
or
\begin{equation}
\left\{ -\frac{d^2}{dx^2} + W^2(x,\alpha(h)) - W'(x,\alpha(h)) -E \right\}
\psi_{h_{min},n}(x) = 0 ~,
\end{equation}
which is exactly the Schr\"odinger equation appearing in ref. \cite{Barclay},
 with eigenenergies  given by eq. (\ref{energy}).  The elegant correspondence
 that exists between potential algebra and supersymmetric quantum mechanics for
 shape invariant potentials is further described in ref. \cite{Chat}.

For a more general case, we assume $R(a_0)=\sum_{j=1}^\infty R_j a_0^j$. In
this case
\begin{equation}
g(h)=\sum_{j=1}^\infty \frac{R_j}{1-e^{jp}}e^{-jph}~~,
\end{equation}
and therefore one gets

\begin{eqnarray}
a^2(h) \eq g(h-n-1)-g(h-1) \nonumber \\
\eq \sum_{j=1}^\infty \alpha_j (h) \frac{1-q^{jn}}{1-q^j}~~,
\end{eqnarray}
where $\alpha_j (h) = R_j e^{-j(h-1)}$.
These results agree with those obtained in ref. \cite{Barclay}.

\subsection{Cyclic Potentials}
Let us consider a particular change of parameters given by the following
cycle (or chain):
\be
\label{chain}
a_0 ,~  a_1=f(a_0) ,~  a_2 = f(a_1)~ ,{\ldots},~
a_{k-1}=f(a_{k-2}) ,~ a_k=f(a_{k-1})=a_0 ,
\ee
and choose $R(a_i)= a_i \equiv \omega_i$.
This choice generates cyclic potentials studied in ref. \cite{Sukhatme}.

Cyclic potentials form a series of shape invariant potentials; the series
 repeats
after a cycle of $k$ iterations. In fig. \ref{cyclic3} we show  the
first potential $V(x,a_0)$ from a $3$-chain ($k=3$) of cyclic potentials,
corresponding to $\omega_0=0.15,\omega_1=0.25,\omega_2=0.60$.
\begin{figure}[ht]
    \centering
    \epsfig{file=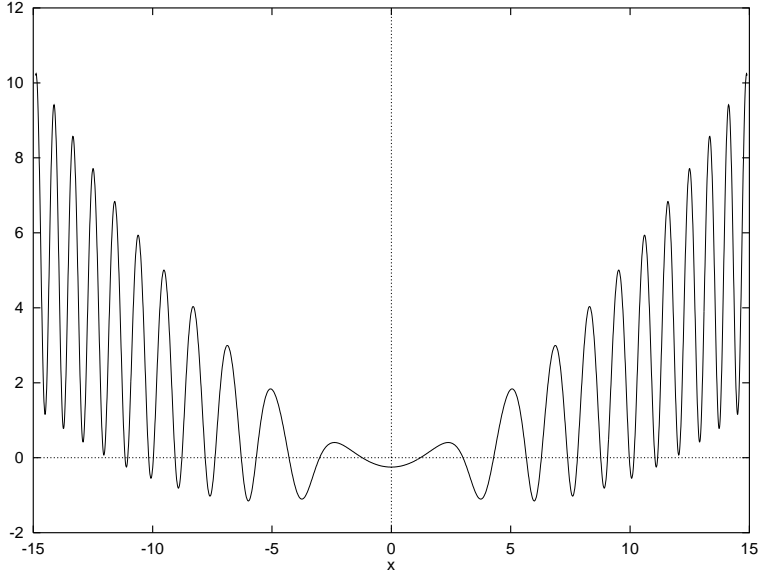, width=3in, angle=-90}
    \caption{First potential $V(x,a_0)$ from a $3$-chain ($k=3$).}
    \label{cyclic3}
\end{figure}

Such potentials have an infinite number of
periodically spaced eigenvalues. More precisely, the level spacings are
given by
$
\omega_0,\omega_1,\ldots,\omega_{k-1},
\omega_0,\omega_1,\ldots,\omega_{k-1},\omega_0,\omega_1,\ldots~,
$

In order to generate the change of parameters (\ref{chain}) the function $f$
should satisfy
$f(f({\ldots} f(x){\ldots} )) \equiv f^k(x) = x$.
The projective map
\be \label{proj}
f(y)=\frac{\alpha y+\beta}{\gamma y+\delta}~~,
\ee
with specific constraints on the parameters $\alpha,\beta,\gamma,\delta$,
 satisfies such a
condition \cite{Sukhatme}.

The next step is to identify the Lie algebra behind this model.
For this, we need to find the function $\chi$ satisfying the equation
\be
\label{cyclic-chi}
\chi(z+p)=f(\chi(z)) \equiv \frac{\alpha \chi(z)+\beta}{\gamma\chi(z)+\delta}~.
\ee
It is a difference equation and its general solution is given by
\be
\label{chi-k}
\chi(z) =
 \frac{(\lambda_1-\delta)\lambda_1^{z/p}+(\lambda_2-\delta)\lambda_2^{z/p}B(z)}
      { \gamma\left[\lambda_1^{z/p}+\lambda_2^{z/p}B(z)\right]}~,
\ee
where $\lambda_{1,2}$ are solutions of the equation
$(x-\alpha)(x-\delta)-\beta\gamma =0$~. For simplicity $B(z)$ can be chosen to
 be an arbitrary constant.
Plugging this expression in eqs. (\ref{lie}) we obtain:
\br
\label{lie-cyclic}
   \lbrack J_3, J_{\pm} \rbrack \eq \pm \;J_{\pm}~~ ;\nonu \\
   \lbrack J_{+},J_{-} \rbrack \eq \xi(J_3)
   \equiv - \frac{1}{c} \frac{A(\lambda_1-\delta)\lambda_1^{-J_3}
     +B(\lambda_2-\delta)\lambda_2^{-J_3}}
{A\lambda_1^{-J_3}+B\lambda_2^{-J_3}}~.
\er

Applying our standard procedure to find the spectrum of the
Hamiltonian $H_-=J_+J_-$ we find that the ground state is at zero energy;
the next $(k-1)$ eigenvalues
are
$
E_l = \sum_{j=0}^{l} \omega_j~,~l=0,1,\ldots,(k-2)~,
$
and all other eigenvalues are obtained by adding arbitrary multiples
of the quantity
\mbox{$\Omega_k \equiv \omega_0+\omega_1+\cdots+\omega_{k-1}$}.
This result is in complete agreement with \cite{Sukhatme}.

\subsection{Natanzon Potentials}

In sec. 2, we noted  that for SIP's with translationally related parameters
$(i.e. ~a_1=a_0+1)$, the shape invariance condition helps in closing the algebra
to the familiar $so(3)$ or $so(2,1)$, provided the
$R(a_0)$ was linear in $a_0$ \cite{Gangopadhyaya_proc}. Several SIP's belong to
 this category; among
them are the Morse, Scarf~ I, Scarf ~II, and generalized P\"oschl-Teller
potentials. However, there are many important SIP's (e.g., Coulomb), whose
associated $R(a_0)$'s are not linear in $a_0$.  Our method of the previous
section would lead to deformed potential algebras for these systems. While we
now know how to get representations of such algebras, in this section we shall
 take a different approach. We choose to generalize the {\it structure} of
 operators $J_\pm$ such that their {\it algebra} still remains linear. In fact,
 in this section, we generate shape invariant potentials from an underlying
 potential algebra instead of showing algebraic structure hidden in a shape
 invariant system.

Alhassid et al.\cite{Alhassid} had shown that the algebra associated with the
{\it general} potential of the Natanzon class is $so(2,2)$. The Schr\"{o}dinger
 equation for these potentials  reduce in general to the hypergeometric
 equation. We show below that a further constraint generates all SIP's with
 translational change of parameters. For the sake of completeness we will
 briefly examine the properties of $so(2,2)$ algebra in this section, and show
 its connection to the
Natanzon potentials \cite{Natanzon}. We then conjecture an additional constraint
 that would render them shape invariant. We find that this conjecture indeed
 produces all known SIP of the translational type. We shall find in fact that
 the subset of
Natanzon potentials associated with the translational (additive) SIP's has
the simpler $so(2,1)$ algebra.

We begin by describing Alhassid et al.'s realization of the $so(2,2)$ algebra in
terms of differential operators. For consistency, we use the formalism and
the notations of refs. \cite{Alhassid}.

The differential realization can be written explicitly as
\br
A_{\pm} \equiv A_1 \pm A_2  &=& {1\over 2}~e^{\pm i(\phi+\theta)}
     \left[ \mp \frac{\partial}{\partial \chi} +
     \tanh \chi \left(-i \partial_\phi \right) +
     \coth \chi \left(-i \partial_\theta \right) \right];
\label{try}
\\
A_3&=& -{i \over 2} \left(\partial_\phi + \partial_\theta \right);
\nn \\
B_{\pm} \equiv B_1 \pm B_2  &=& {1\over 2}~e^{\pm i(\phi-\theta)}
     \left[ \mp \frac{\partial}{\partial \chi} +
     \tanh \chi \left(-i \partial_\phi \right) +
     \coth \chi \left(+i \partial_\theta \right) \right];
\nn \\
B_3&=& -{i \over 2} \left(\partial_\phi - \partial_\theta \right)~ .
\nn
\er
The $so(2,1)$ algebra obeyed by these operators is
\br
[A_3, A_\pm] = \pm A_\pm, && [A_+,A_-] = -2A_3~, \nn
\er
and a similar one for the $B$'s. The Casimir operator $C_2$ is given by
\br
C_2 &=& 2~\left(A_3^2 - A_+A_- - A_3 \right)
     + 2~\left(B_3^2 - B_+B_- - B_3 \right) \nonumber \\
    &=&
     \left[ \frac{\partial^2}{\partial \chi^2}
     +\left( \tanh \chi + \coth \chi \right)
      \frac{\partial}{\partial \chi}
     + \sech^2\chi    \left(-i \partial_\phi \right)^2
     - \cosech^2 \chi \left(-i \partial_\theta \right)^2
     \right]  .
\label{casimir}
\er
Operators $A_3$, $B_3$ and $C_2$ can be simultaneously diagonalized, and their
actions on their common eigenstate are given by
\br
C_2 |\omega,m_1,m_2 \rangle &=& \omega (\omega+2) ~|\omega,m_1,m_2 \rangle ~;
\nn \\
A_3 |\omega,m_1,m_2 \rangle &=& m_1 ~|\omega,m_1,m_2 \rangle ~;
\nn \\
B_3 |\omega,m_1,m_2 \rangle &=& m_2 ~|\omega,m_1,m_2 \rangle~.
\label{action}
\er
(It is worth mentioning at this point that the Casimir operator given above
is indeed self-adjoint with respect to a measure
$\sinh\chi\cosh\chi d\chi d\phi d\theta$.)

Now we shall briefly describe a general Natanzon potential and show its
 connection to the above Casimir operator. A general Natanzon potential $U(r)$
 is implicitly defined by \cite{Natanzon}
\be
U[z(r)] = \frac{-f   z(1-z) + h_0   (1-z) + h_1   z}{Q(z)}
-{1\over 2} \left\{ z,r \right\}~,
\label{Natanzon_pot}
\ee
with $Q(z)$ quadratic in $z$: $Q(z)=a z^2 +b_0 z + c_0 = a (1-z)^2 -b_1 (1-z)+
 c_1$ and
$f, h_0, h_1, a, b_0, b_1, c_0, c_1$ are constants. The Schwarzian derivative
$\left\{ z,r \right\}$ is defined by
\be
\left\{ z,r \right\} \equiv \frac{d^3z/dr^3}{dz/dr} -{3 \over 2}
\left[ \frac{d^2z/dr^2}{dz/dr} \right]^2 ~.
\label{Schwarzian}
\ee
The relationship between variables $z$ ($0<z<1$) and $r$ is
implicitly given by
\be
\left( \frac{dz}{dr} \right) = \frac{2z(1-z)}{\sqrt{Q(z)}}~.
\label{z,r}
\ee

To connect the Casimir operator $C_2$ of the $so(2,2)$ algebra [eq.
 (\ref{casimir})] to the general Natanzon potential, we perform a similarity
 transformation on $C_2$ by a function $F$ and then follow that up by an
 appropriate change of variable $\chi =g(r)$. It has been shown
 \cite{Gangopadhyaya_pra} that to turn $C_2$ into the form of a Schr\"odinger
 Hamiltonian, one needs to choose $F \sim  \left( \frac{\sinh
 (2g)}{g'}\right)^{1\over 2}~$, and choose $z=\tanh^2g$.
Then \br
U(z(r))&=&\frac{E~Q+[-{7 \over 4}+{5 \over 2}z-
{7 \over 4} z^2]-z(1-z) \left(-i\partial_\phi \right)^2
+(1-z)\left(-i\partial_\theta \right)^2 }{Q} -{1\over 2} \left\{z,r\right\}
\nn\\
&&\nn\\
&=&\left[
-\left( aE-{7 \over 4}+\left(-i\partial_\phi \right)^2 \right)z(1-z)
+ \left(c_0 E -{7 \over 4} + \left(-i\partial_\theta \right)^2 \right) (1-z)
\right. \nn \\
&& \left. +\left( (a+b_0+c_0) E -1 \right) \right] / Q(z)
~-{1\over 2} \left\{z,r\right\}
~~.
\label{Natanzon2}
\er
We have used
$$g'=\frac{dg}{dr}=\frac{dg}{dz}\frac{dz}{dr}= \frac{1}{2 \sqrt{z} (1-z)} ~
\frac{2z(1-z)}{Q}= \sqrt{\frac{z}{Q}}~~,$$
 $g=\tanh^{-1}\sqrt{z}$ and $\frac{dz}{dr}$ from eq. (\ref{z,r}).  Now, with the
 following identification
\br
f   &=& aE-{7 \over 4}+\left(-i\partial_\phi \right)^2 , \nn\\
h_0 &=& c_0 E -{7 \over 4} + \left(-i\partial_\theta \right)^2 ,\nn\\
h_1 &=& (a+b_0+c_0)E  -1 ~~,
\er
the potential of eq. (\ref{Natanzon2}) indeed has the form of a general
Natanzon potential [eq. (\ref{Natanzon_pot})].
Further details are shown in ref. \cite{Gangopadhyaya_pra}. Finally, we are
 ready to explicitly demonstrate the connection between the Natanzon potential
 algebra and shape invariant potentials of supersymmetric quantum mechanics.

We now note that the similarity transformation can be rewritten: using
 $g=\tanh^{-1}\sqrt{z}$,
$g'=\sqrt{\frac{z}{Q}}$ and eq. (\ref{z,r}), we find $\left( \frac{\sinh
 (2g)}{g'}\right)=
\frac{z}{z'}$.
At this point we go back to the operators $A_\pm$
[eq. (\ref{try})]
and ask how they transform under the similarity transformation given by
$F \sim \left( \frac{\sinh (2g)}{g'}\right)^{1\over 2} \sim \sqrt{z\over{z'}}$.
This transformation carries operators $A_\pm$ to
\be
A_\pm \longrightarrow {\tilde A}_\pm =
     {{e^{\pm i (\phi+\theta)} }\over 2}
\left[
     \mp \left( \frac{d}{d\chi}  + \frac{1}{2z'} \frac{dz'}{d\chi}
           - \frac{1}{2z} \frac{dz}{d\chi} \right)
           +\tanh\chi ~(-i\partial_\phi)+\coth\chi ~(-i\partial_\theta)
\right]~.
\ee
Except for the expression $\left(\frac{1}{2z'} \frac{dz'}{d\chi}-\frac{1}{2z}
\frac{dz}{d\chi} \right)$, this looks very much like eq. (\ref{try}), which are
 in fact $A_\pm$ of the shape invariant P\"{o}schl-Teller
 potential\cite{Cooper}.
Thus, if $\left(\frac{1}{2z'} \frac{dz'}{d\chi}-\frac{1}{2z}
\frac{dz}{d\chi} \right)$ were to be a linear combination of $\tanh\chi$
and $\coth\chi$, operators ${\tilde A}_\pm$ could be cast in a form similar to
 the
operators $A_\pm$ of eq. (\ref{try}), and we would get $A_\pm$'s that generate
 shape
invariant Hamiltonians.

Hence to get shape invariant potentials, we require,
\be
\left( \frac{1}{2z'}
\frac{dz'}{d\chi}-\frac{1}{2z} \frac{dz}{d\chi} \right) =\alpha\tanh\chi +
\beta\coth\chi.
\ee
This leads to
\be
z'=z^{1+\beta} \dot (1-z)^{-\alpha-\beta},
\label{stipulation}
\ee
which is the second constraint on the relationship between variables $z$ and
 $r$. Since these variables are already constrained by eq. (\ref{z,r}), only a
 handful of solutions would be compatible with both restrictions. The $z(r)$'s
 that are compatible with both
eqs. (\ref{z,r}) and (\ref{stipulation})are given by
\be
z^{1+\beta} \dot (1-z)^{-\alpha-\beta}=\frac{2z(1-z)}{\sqrt{Q(z)}},
\label{constraint}
\ee
where $Q(z)$ is a quadratic function of $z$. After some computation, we find
 that there is only a finite number of values of $\alpha$, $\beta$ which satisfy
 eq. (\ref{constraint}). These values are listed in Table 1, and they exhaust
 all known shape invariant potentials that lead
to the hypergeometric equation.

Furthermore, while the potential algebra of a general Natanzon system is
 $so(2,2)$, and requires two sets of raising and lowering operators $A_{\pm}$
 and $B_{\pm}$, all translational shape invariant potentials need only one such
 set. For all SIPs of Table 4.1 of ref. \cite{Cooper}, one finds that all
 partner potentials are connected by change of just one independent parameter
 (although other parameters which don't change are also present.) Thus there is
 a series of potentials that only differ in one
parameter. From the potential algebra perspective, all these potentials  differ
 only by the eigenvalue of an operator that is a linear combination of $A_3$ and
 $B_3$, and all are characterized by a common eigenvalue of
$C_2$. Thus, these shape invariant potentials can be associated with a
$so(2,1)$ potential algebra generated by operators $A_+$, $A_-$ and the same
 linear combination of $A_3$ and $B_3$.

Conclusion: In this paper, we have explored the reasons underlying the
 integrability of shape invariant Hamiltonians and shown that such systems
 naturally admit an algebraic structure known as potential algebra. We have
 derived these algebras for shape invariant systems with translational and
 scaling type change of parameters, as well as for cyclic potentials. In
 general, one finds deformations of the $so(2,1)$ Lie algebra. Our approach
 links the group theoretic and supersymmetric quantum mechanics approaches for
 treating shape invariant potentials.

A.G. acknowledges a research leave and a grant from Loyola University Chicago
 which made his involvement in this work possible. Partial financial support
 from the U.S. Department of Energy is gratefully acknowledged.

\begin{center}
\begin{tabular}{||c|c|l|l|l||}
\hline&&&&\\
{\bf $\alpha$} & {\bf $\beta$}  &  {\bf $z(r)$} &    {\bf Superpotential} &
{\bf Potential}\\
&&&&\\
\hline&&&&\\
 $0$      &  $0$    & $z=e^{-r}$   &$ \tilde{m}_1 \coth \frac{r}{2} +
\tilde{m}_2$      & Eckart\\
     &         &              &                        &\\
\hline&&&&\\
 $0$      &$ -{1\over 2}$     & $z=\sin^2\frac{r}{2}$  &$ \tilde{m}_1 \cosec
\,r +  \tilde{m}_2
\cot r $
                                        &        Gen.  P\"oschl-Teller\\
     &         &              &                   &   trigonometric\\
     &         &              &                        &\\
\hline&&&&\\
 $0$      & $-1$         &$z=1-e^{-r}$            &$ \tilde{m}_1 \coth
\frac{r}{2} + \tilde{m}_2$
  &      Eckart \\
     &         &              &                        &\\
\hline&&&&\\
$-{1\over 2}$ &  $0$     &$z=\sech^2\frac{r}{2}$  &$ \tilde{m}_1 \cosech
\,r +  \tilde{m}_2
\coth r $
 & P\"oschl-Teller II \\
     &         &              &                        &\\
\hline&&&&\\
$-{1\over 2}$  & $-{1\over 2} $ &$z=\tanh^2\frac{r}{2} $&$ \tilde{m}_1 \tanh
\frac{r}{2}
+  \tilde{m}_2 \coth \frac{r}{2} $     &    Gen. P\"oschl-Teller \\
     &         &              &                        &\\
\hline&&&&\\
$-1$      &  $0$         &$z=1+\tanh \frac{r}{2}$      & $ \tilde{m}_1 \tanh
\frac{r}{2} + \tilde{m}_2$
&        Rosen ~Morse\\
     &         &              &                        &\\
\hline
\end{tabular}

\vspace*{.25in}
{Table 1: Shows all allowed value of $\alpha,~\beta$
and the superpotentials that they generate.}
\end{center}
\end{document}